\documentclass[useAMS,usenatbib,usegraphicx]{mn2e}

\title[The ultraviolet colour of globular clusters in M31] 
  {The ultraviolet colour of globular clusters in M31: a core density effect?}
\author[M. Peacock et al.]
{Mark B. Peacock$^{1}$\thanks{E-mail:mpeacock@msu.edu (MBP)},
Thomas J. Maccarone$^{2}$, Andrea Dieball$^{2}$, Christian Knigge$^{2}$ \\
$^{1}$Department of Physics and Astronomy, Michigan State University, East Lansing, MI 48824, USA\\
$^{2}$School of Physics and Astronomy, University of Southampton, Southampton, SO17 1BJ, UK}

\begin{document}

\date{Released 2010 Xxxxx XX}

\pagerange{\pageref{firstpage}--\pageref{lastpage}} \pubyear{2010}

\maketitle

\label{firstpage}

\begin{abstract}
\label{sec:fuv:abstract}

We investigate the effect of stellar density on the ultraviolet (UV) emission from M31's globular clusters (GCs). Published far-UV (FUV) and near-UV (NUV) colours from \textit{Galaxy Evolution and Explorer (GALEX)} observations are used as a probe into the temperature of the horizontal branch (HB) stars in these clusters. From these data, we demonstrate a significant relationship between the core density of a cluster and its FUV-NUV colour, with dense clusters having bluer ultraviolet colours. These results are consistent with a population of (FUV bright) extreme-HB (EHB) stars, the production of which is related to the stellar density in the clusters. Such a relationship may be expected if the formation of EHB stars is enhanced in dense clusters due to dynamical interactions. We also consider the contribution of low mass X-ray binaries (LMXBs) to the integrated FUV luminosity of a cluster. We note that two of the three metal rich clusters, identified by \citet{Rey07} as having a FUV excess, are known to host LMXBs in outburst. Considering the FUV luminosity of Galactic LMXBs, we suggest that a single LMXB is unlikely to produce more than 10$\%$ of the observed FUV luminosity of clusters that contain a significant population of blue-HB stars. 

\end{abstract}

\begin{keywords}
globular clusters: general - binaries: general - X-rays: binaries
\end{keywords}

\section{Introduction}
\label{sec:fuv:intro}

Horizontal branch (HB) stars are core helium-burning stars that have evolved off the red giant branch (RGB). In globular clusters (GCs) they are so named because of their appearance on optical colour-magnitude diagrams, where they trace a horizontal path blueward of the RGB. These stars are thought to have similar core masses of $\sim$0.5$M_{\odot}$. Their location along the HB is therefore determined primarily by the mass and opacity of their thin stellar envelopes. To explain the observed HB stars in GCs, it is thought that substantial mass loss must occur during the star's RGB phase \citep[e.g.][]{Rood73}. The HB morphologies of different Galactic GCs are known to vary significantly. This suggests that the cluster properties have a significant effect on the evolution of stars onto the HB. The metallicity of a GC has long been proposed as the `first parameter' related to the morphology of a GC's HB, with metal poor clusters often having bluer HB stars than metal rich GCs \citep[e.g.][]{Sandage60,Dorman95}. However, the metallicity of a cluster alone is insufficient to explain the HB morphology in all Galactic GCs. Firstly, the clusters which host the bluest HBs actually have intermediate metallicities \citep[e.g.][]{OConnell99}. Also, some clusters are observed to have very different HB morphologies, but similar metallicities \citep[e.g. NGC~288 and NGC~362:][]{Bellazzini01}. This has led to the notorious search for a `second parameter' to describe HB morphology. Many different parameters have been proposed as a second parameter, all of which may play some role. These include the cluster's age \citep[e.g.][]{Demarque88,Lee94}, its helium abundance \citep[e.g.][]{Sweigart97,Lee05}, the stellar core rotation \citep[e.g.][]{Mengel76,Peterson85}, mass loss on the RGB \citep[e.g.][]{Rood73} and cluster core density \citep{Buonanno97}. Despite decades of research, the relative effects of these different parameters are yet to be fully understood \citep[for a discussion of proposed second parameters, see e.g. ][]{Fusi_Pecci97,Catelan09}. 

The bluest HB stars observed in GCs are the `blue tail' or `extreme-HB' (EHB) stars. These are analogues to the subdwarf~B (sdB) stars observed in the Galactic field and are considered here to to be those stars with T$_{eff}>$20,000K \citep{Brown01}. These stars have become increasingly important as the leading contenders in explaining the `ultraviolet excess' (UVX) observed in elliptical galaxies \citep[e.g.][]{Code69,Dorman95,OConnell99}. Because of their high effective temperatures, they are relatively faint at optical wavelengths but make an important, often dominating, contribution to the FUV emission from old stellar populations. EHB stars are observed in both metal rich and metal poor clusters \citep[e.g.][]{DCruz96,Rich97} and it is likely that the metallicity may have little direct effect on the formation of these stars \citep[e.g.][]{Heber86,Dorman93,OConnell99}. Due to their very low envelope masses \citep[$\la$0.02M$_{\odot}$][]{Heber86}, it is likely that their formation and colour is more strongly related to the efficiency of mass loss mechanisms than the metallicity. At the hottest and optically faintest end of the EHB population, some Galactic clusters are observed to host a population of `blue hook' stars \citep[BHk;][]{Brown01,Dieball09}. These stars have similar UV colours to EHB stars, but are slightly fainter, lying just below the zero age horizontal branch population in UV colour magnitude diagrams (CMDs). Although BHk stars appear to be relatively rare, if a cluster hosts a significant population of these stars, then they are likely to make a contribution to the FUV luminosity of the cluster. 

In this study, we investigate the effect of a cluster's core density on its HB population. A relationship has previously been proposed between core density and the extent of the HB population in the Milky Way's GCs \citep{Fusi_Pecci93,Buonanno97}. There are several reasons to suspect that dense stellar environments will enhance the formation of EHB stars. Firstly, mass loss may be enhanced through tidal interactions in close encounters between stars. Close encounters may also increase mixing in RGB stars and lead to helium enrichment, which will make the HB stars bluer \citep{Suda07}. It has also been proposed that EHB stars may be formed in binary systems either via white dwarf (WD)-WD mergers or from mass loss through either Roche lobe overflow or common envelope mechanisms \citep{Han02,Han03,Han07}. This theory is supported by observations of Galactic sdB stars, a large fraction of which are found to be binary systems \citep[e.g.][]{Maxted01,Reed04}. Interestingly, current observations suggest that the binary fraction among EHB stars may be much lower in GCs than in the field \citep{Moni_Bidin09}. This may be due to the relative contributions of different formation mechanisms evolving with time, with WD-WD mergers becoming increasingly import for old dense stellar populations, such as GCs \citep{Moni_Bidin08,Han08}. If EHB stars do have a binary origin, then a relationship between the core density of a cluster and the population of EHB stars it hosts may be expected. This is because of dynamical formation of these binary systems in the cores of these clusters. 

Only a few of M31's GCs have CMDs which allow direct observation of their HB stars, and these are generally incomplete \citep[e.g.][]{Rich05}. Even in the Milky Way's GCs, studying EHB/BHk stars based on optical observations is difficult due to the severe crowding of sources and because these stars are optically faint. They are better identified in FUV CMDs, although such observations of Galactic clusters are severely affected by extinction. This means that only a few Galactic GCs have been studied from UV CMDs \citep[see e.g.][]{Brown01,Knigge02,Dieball05,Dieball10}. In the absence of direct stellar counts for M31's clusters, we consider their integrated FUV and NUV colours. This photometry provides a probe into the hot stellar populations in these GCs. Main sequence and RGB stars are too cool to have significant emission at these wavelengths and the clusters' integrated FUV and NUV magnitudes are likely to be dominated by objects hotter than $\sim$10,000K. If a cluster hosts a population of blue-HB (BHB), EHB or BHk stars, then these will likely dominate the cluster luminosity at these wavelengths. In the Milky Way, the integrated FUV-V colour is found to correlate with the HB morphology of the cluster \citep[e.g. fig. 3 of ][]{Catelan09}. 

Recent \textit{GALEX} FUV and NUV observations of M31 have been presented by \citet{Rey05,Rey07}. In their work, the expected relationship between the clusters metallicity and both its FUV-V and NUV-V colour were shown. However, there is significant scatter in the FUV relationship. This confirms the need for a second parameter to describe the FUV luminosity of clusters in M31. \citet{Rey07} also identify three metal rich clusters which have a FUV excess, compared with both stellar models and other metal rich clusters in the galaxy. To help explain their observations, they consider the effects of age and helium abundance on the FUV luminosities of M31's GCs. Helium abundance is also used by \citet{Sohn06} and \citet{Kaviraj07a} to explain the FUV observations of M87's GCs. In this study, we consider correlations between the luminosity and core density of a GC and its FUV properties. 

\section{M31 globular cluster data}
\label{sec:fuv:data}

In the following analysis, we use data from the recent catalogue of the M31's GCs from \citet{Peacock10}. This catalogue provides updated classifications, \textit{ugriz} and K-band photometry and structural parameters for objects listed in the Revised Bologna Catalogue of GCs in M31 \citep[RBC:][]{Galleti04}. We use this catalogue to identify 416 confirmed old clusters in the galaxy, of which 213 have structural parameters. The clusters with estimated structural parameters are spatially limited to the disk region of M31 (minus the very central region) and cover the entire GC luminosity function \citep[although they are thought to be relatively unreliable for the faintest clusters, as discussed in][]{Peacock09}. As the primary aim of this work is to investigate the effect of stellar density on the UV properties of the clusters, we restrict our primary sample to those clusters with structural parameters. 

We combine this catalogue with the spectroscopic metallicities collated by \citet{Fan08} from the studies of \citet{Huchra91,Barmby00,Perrett02}. Where these metallicities are available from more than one study, the values from the larger and more recent catalogue of \citet{Perrett02} are chosen over \citet{Barmby00}. We reject the cluster B235 because of a large discrepancy ($\Delta$[Fe/H]=0.6) between the metallicities available from \citet{Barmby00} and \citet{Perrett02}. We also remove those clusters with large errors ($\sigma_{\rm{[Fe/H]}}>$0.6) on their metallicities (this removes the clusters B214, B229, B352 and BA11 from our analysis). This gives spectroscopically estimated metallicities for 219/416 confirmed old clusters in our catalogue. \citet{Fan08} also present reddening estimates for these clusters based on this spectroscopy and on photometry from the RBC. We use these estimates of $E(B-V)$ to deredden our photometry using the extinction curves of \citet{Cardelli89}. The relative extinction in the different bands were evaluated from these curves by \citet{Schlegel98} (for the \textit{ugriz} filters) and by \citet{Rey07} (for the \textit{GALEX} FUV, $R_{\rm{FUV}}$=8.16 and NUV, $R_{\rm{NUV}}$=8.90). These values are relatively uncertain. This is because the reddening curve in the UV is known to vary along different lines of sight in the Milky Way. It is also not guaranteed that M31 has the same extinction curve as the Milky Way, although \citet{Barmby00} found no evidence of significant differences in the optical bands. To minimise the effects of this on our UV colours, we limit our analysis to the clusters with relatively low extinction \citep[$E(B-V)<$0.16, consistent with the limit adopted by][]{Rey07}. 

\begin{figure}
 \centering
 \includegraphics[height=84mm,angle=270]{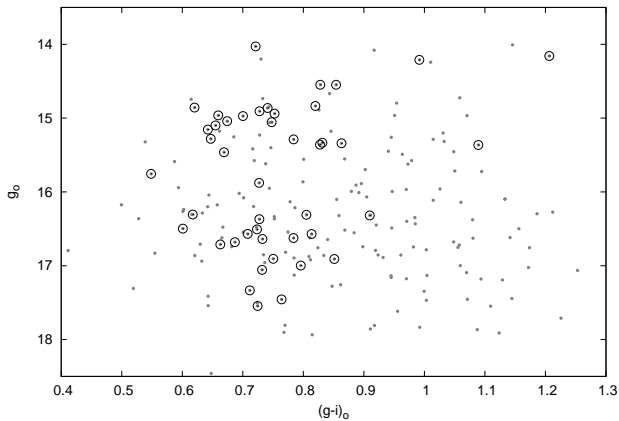}
 \caption{Optical colour magnitude diagram for all M31 GCs with data from \citet{Peacock10} and E(B-V)$<$0.16 from \citet{Fan08}. Small grey points indicate all GCs, while open black circles indicate those clusters detected in the \textit{GALEX} FUV observations.}
 \label{fig:mag_colour}
\end{figure}

\begin{figure*}
 \includegraphics[width=100mm,angle=270]{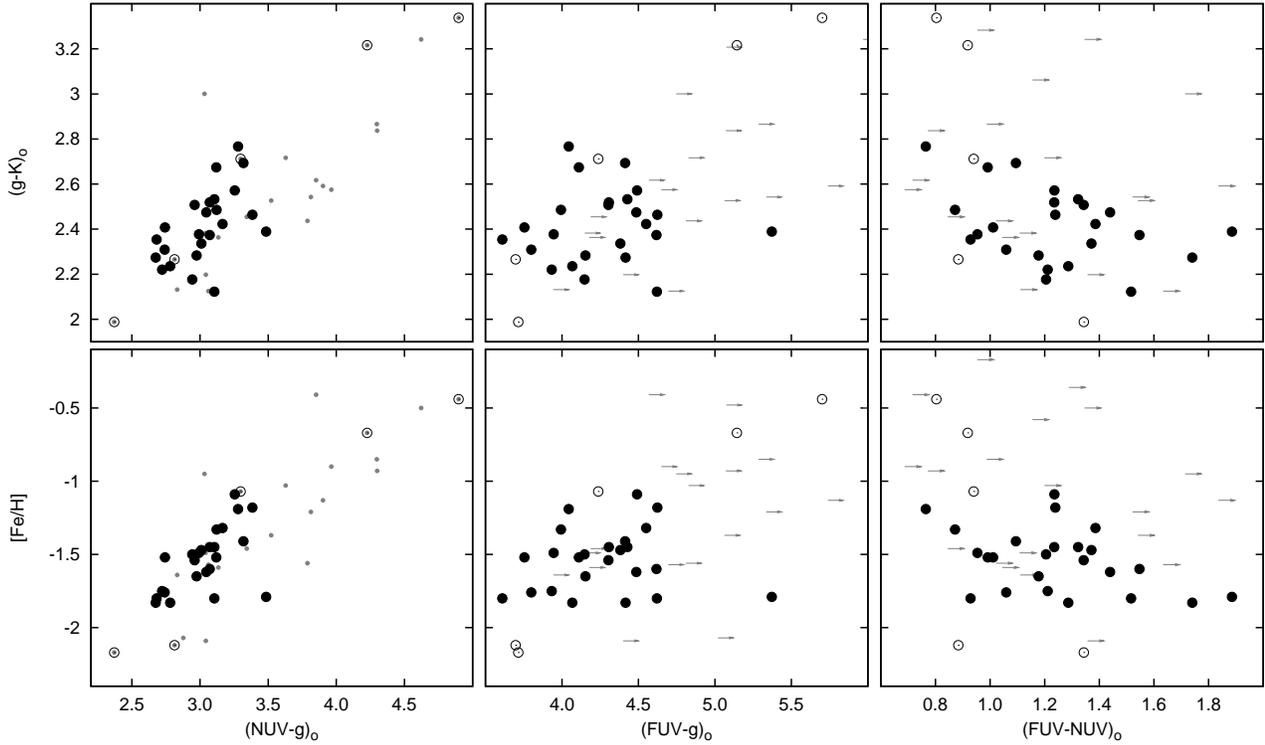}
 \caption{The ultraviolet colours of M31's GCs as a function of both \textit{g}-K colour (\textit{top}) and metallicity (\textit{bottom}). Colours are dereddened using the reddening estimates of \citet{Fan08}. The grey points/arrows show those clusters detected in the NUV but not the FUV. Filled circles indicate clusters with intermediate metallicities (-1.8$<$[Fe/H]$<$-1.2).}
 \label{fig:UV_Z}
\end{figure*}

\begin{figure*}
 \includegraphics[width=100mm,angle=270]{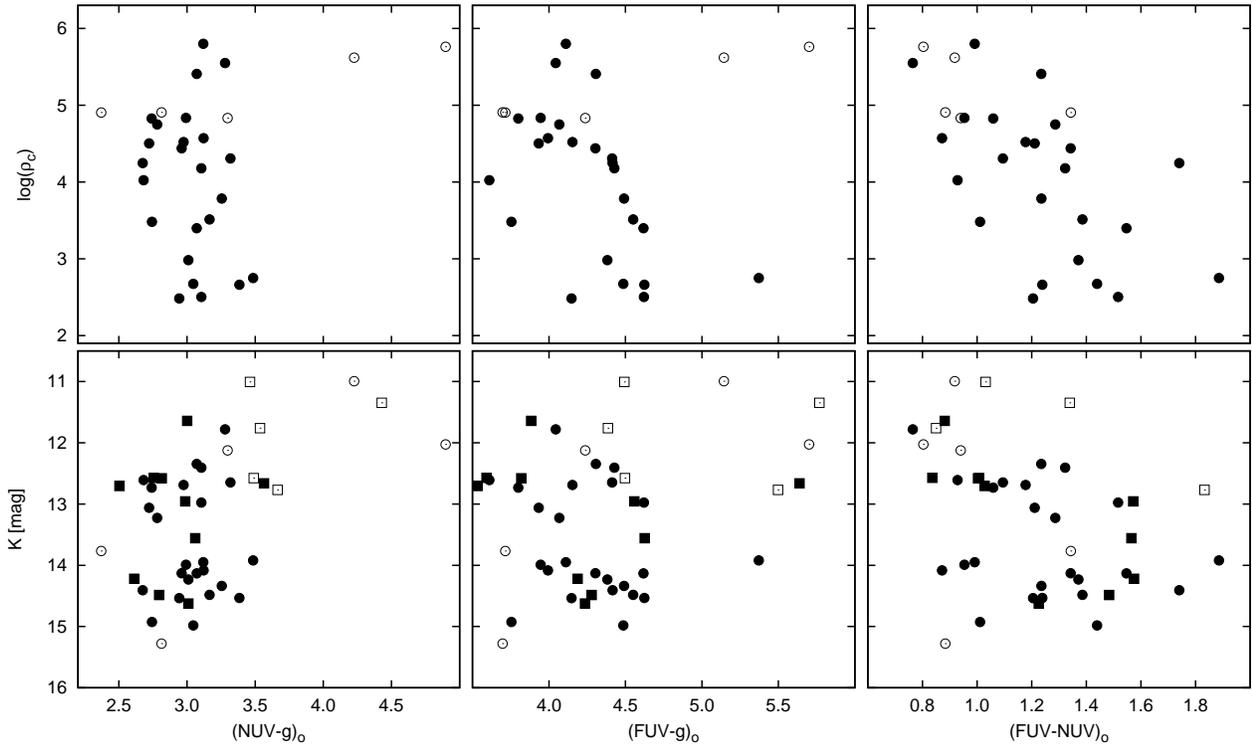}
 \caption{The ultraviolet colours of M31's GCs as a function of core density (\textit{top}) and K-band luminosity (\textit{bottom}). The bottom figures include 15 additional clusters, not covered by the survey of \citet{Peacock09} but, whose K-band luminosities are available from 2MASS observations (squares). As in figure \ref{fig:UV_Z}, filled symbols indicate intermediate metallicity clusters. }
 \label{fig:UV_lumin_rho0}
\end{figure*}

The FUV and NUV magnitudes of these clusters are taken from \citet{Rey05,Rey07}. These data provide integrated FUV and NUV magnitudes for 104 and 210 of the 416 confirmed old clusters in \citet{Peacock10}. For full details of these UV observations, we refer the reader to \citet{Rey05,Rey07}. The \textit{GALEX} FUV (1344-1786$\rm{\AA}$) and NUV (1771-2831$\rm{\AA}$) filters have an effective wavelength of 1528$\rm{\AA}$ and 2271$\rm{\AA}$, respectively \citep{Morrissey05}. The photometry is on the standard \textit{GALEX} ABmag photometric system, where: 
\begin{equation}
 \label{eq:fuv_flux}
 \rm{FUV_{AB}} = -2.5\times\rm{log_{10}(F_{FUV}/1.40\times10^{-15}}) + 18.82 
\end{equation}
\begin{equation}
 \label{eq:nuv_flux}
 \rm{NUV_{AB}} = -2.5\times\rm{log_{10}(F_{NUV}/2.06\times10^{-16}}) + 20.08
\end{equation}
Here, F is the flux of the source (erg $\rm{sec^{-1}cm^{-2}\AA^{-1}}$). The sample of clusters is magnitude limited to $\sim$22.5 in both the FUV and NUV. As highlighted by figure \ref{fig:mag_colour}, this does not correspond to a homogeneous optical magnitude limit. It can be seen that the blue (metal poor) clusters are detected across the GC luminosity function. However, very few of the red (metal rich) clusters are detected in the FUV. This is discussed by \citet{Rey07} and highlights the effect of metallicity on the HB properties, and hence FUV brightness, of a cluster. This plot also identifies the three metal rich clusters that are detected due to proposed excess FUV emission, as discussed by \citet{Rey07}. 

Our final dataset contains 51 confirmed old clusters which have $E(B-V)<$0.16, NUV and ugriz photometry, estimates of their core densities and spectroscopic metallicities. Of these clusters 29 have FUV photometry.

\section{UV properties of M31 clusters}

\subsection{Metallicity relationships} 

Figure \ref{fig:UV_Z} shows the FUV-\textit{g} and NUV-\textit{g} colours of M31's clusters as a function of metallicity and \textit{g}-K colour. It can be seen that the NUV-\textit{g} colour of a cluster is strongly correlated with its metallicity. Although relatively weak, the FUV-\textit{g} colour is also correlated with the metallicity. Considering the clusters with only upper limits available for their FUV magnitude confirms this relationship, demonstrating a lack of FUV bright, metal rich, clusters. These relationships are to be expected and highlight the effect of metallicity on the HB population in a cluster. This result was already identified by \citet{Rey07}. It is included here for completeness, but for a more detailed discussion, we refer the reader to this work. 

The bottom right panel of figure \ref{fig:UV_Z} shows the metallicity as a function of FUV-NUV colour. It can be seen that the relationship is much less clear in this colour. A Spearman rank test actually shows that \textit{metal rich} clusters appear to be slightly bluer (with number of clusters, N=29, Spearman rank correlation coefficient, $\rho_{s}$=-0.39 and P-value for non-correlation, P=0.03). There is no obvious reason for such a relationship and it should be noted that selection effects, which bias us against FUV faint, metal rich clusters, could result in such a relationship. Also, this relationship is dominated by the two metal rich clusters in our sample. The FUV luminosity of these metal rich clusters may follow a different relationship to metal poor and intermediate metallicity clusters. However, it can be seen from these data that, unlike the UV-optical colour, there is no evidence for the FUV-NUV colour increasing with the metallicity of a cluster. This suggests that the effect of metallicity is much weaker on the UV colour of these clusters.  

The two metal rich clusters in our sample have very blue FUV-NUV colours. This could result from selection effects, as these bias us against all but the FUV brightest metal rich clusters. However, the blue colours of these clusters could indicate that the FUV output of metal rich clusters (which generally have redder horizontal branches and few BHB stars) may be dominated by a population of EHB stars. EHB stars will have similar (or brighter) FUV luminosities to BHB stars. However, they have significantly bluer FUV-NUV colours \citep[as shown by the UV CMD of e.g. NGC~2808 and M80:][]{Brown01,Dieball10}.

\subsection{Stellar density$/$ Luminosity relationships}

As discussed in section \ref{sec:fuv:intro}, one may expect that the density of a cluster could effect its FUV luminosity. Figure \ref{fig:UV_lumin_rho0} shows the FUV-\textit{g}, NUV-\textit{g} and FUV-NUV colours of M31's GCs as a function of K-band luminosity and core density. 

\begin{figure}
 \includegraphics[width=84mm,angle=0]{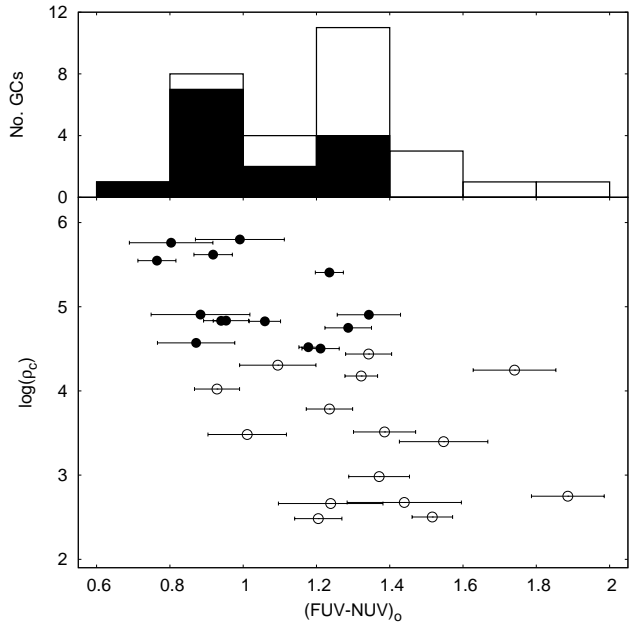}
 \caption{\textit{Bottom panel:} The dereddened FUV-NUV colour of M31's GCs as a function of core density. Solid circles show the densest 50$\%$ of clusters. \textit{Top panel:} compares the FUV-NUV colour of all clusters in our sample (open) with the densest 50$\%$ of clusters in our sample (solid). It can be seen that the denser clusters appear to have bluer UV colours.}
 \label{fig:UV_rho0_limits}
\end{figure}

It can be seen from the top right panel of figure \ref{fig:UV_lumin_rho0} that the cluster core density appears to correlate with the FUV-NUV colour of the clusters. A Spearman rank test demonstrates a significant relationship between these parameters, with higher density clusters having bluer FUV-NUV colours (with N=29, $\rho_{s}$=-0.65 and P=$1.4\times10^{-4}$). Since EHB stars are relatively bright at FUV wavelengths, bluer UV colours could be produced by a larger fraction of EHB stars with respect to BHB stars. The observed relationship could thus be suggestive of a population of EHB stars, the formation of which is related to the core density of the cluster. This relationship is shown in more detail in figure \ref{fig:UV_rho0_limits}. In this figure we have highlighted the densest 50$\%$ of clusters in our sample. It can be seen that the densest group of clusters do appear to be offset to bluer colours. A Kolmogorov-Smirnov test between the high and low density clusters indicates that there is a low probability of 0.8$\%$ that they are drawn from the same population. The lack of dense, FUV faint, clusters can not be explained by selection effects. This is because the denser clusters are also the brighter clusters, and hence generally detected. It can be seen that there is significant scatter in the observed relationship. This is to be expected for two reasons. Firstly, if there is a relationship between the extent of the HB in a cluster and its core density, this is very unlikely to be the only parameter involved. Secondly, the UV colour is an indirect measure of the distribution of HB stars in a cluster. Because of this, other hot populations may weaken genuine trends. 

The bottom panels of figure \ref{fig:UV_lumin_rho0} show the UV colours of the clusters as a function of their total luminosity. To increase the number of clusters studied, we also include in this figure 15 additional clusters, that were not covered by the survey of \citet{Peacock09}. The K-band luminosities of these clusters were taken from the 2MASS survey, as presented by \citet{Galleti04}. A relationship is observed between the total K-band luminosity of a cluster and its UV colour, with the brightest clusters having blue UV colours (N=44, $\rho_{s}$=0.39 and P=0.009). This trend could be due to the same processes as the core density relationship, because of the strong correlation between the density and luminosity of a cluster \citep[as shown in e.g.][]{Peacock09}. However, the relationship may also be suggestive of self enrichment. The greater potential of the most massive clusters may result in them retaining more of the helium they produce. Since higher helium fractions are known to produce bluer HB stars, this could potentially produce the observed trend. Because of the relatively close relationship between the luminosity and density of a cluster, and the weakness of the observed correlations, it is very difficult to distinguish between the relative effects of these parameters. 

As discussed previously, the metallicity of a cluster is expected (and observed) to effect its UV luminosity. For this reason, we also consider a sub-sample of intermediate metallicity clusters (with -1.8$<$[Fe/H]$<$-1.2; highlighted as filled symbols in figures \ref{fig:UV_Z} and \ref{fig:UV_lumin_rho0}). For the Galactic GCs, this method is found to highlight the effects of other parameters on HB stars \citep[since it reduces the influence of metallicity; e.g.][]{Recio_Blanco06}. In this sample, we observe the same core density and luminosity correlations found previously. However, these have a lower significance. This is likely due to the smaller sample size combined with the fact that some of the brightest and densest clusters (in which such an effect should be most evident) lie outside this metallicity range. 

\begin{figure}
 \centering
 \includegraphics[height=66mm,angle=270]{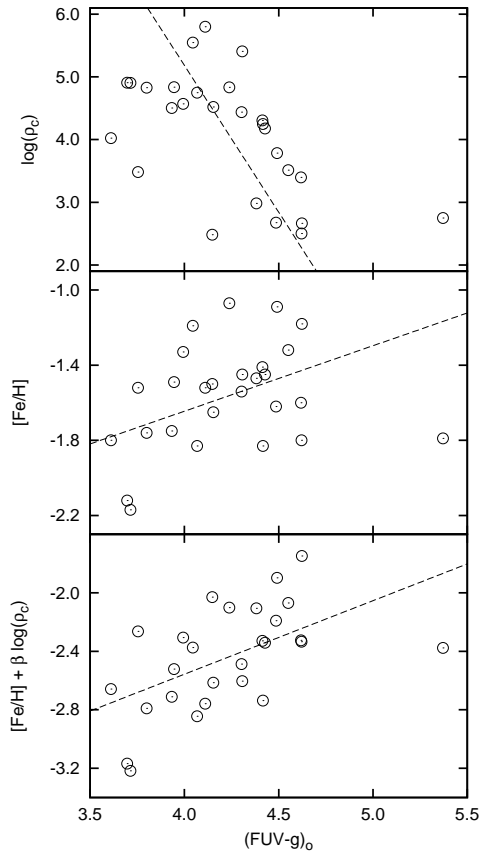}
 \caption{FUV-\textit{g} colour as a function of core density (\textit{top}) and metallicity (\textit{middle}) for the metal-poor clusters (with [Fe/H]$<$-0.8). The \textit{bottom} panel shows the FUV-\textit{g} - metallicity relationship, detrended for the derived core density relationship. }
 \label{fig:fit_FUVg_Z_rho0}
\end{figure}

No significant trends are observed between the clusters K-band luminosities, or their core densities, and their FUV-\textit{g} or NUV-\textit{g} colours. This suggests that the UV-optical colours of these clusters are dominated by other effects, such as the cluster metallicity. However, since the effect of metallicity, on the FUV luminosity of a cluster, is strongest between the metal rich and metal poor clusters (as demonstrated by the low number of metal rich clusters detected in our data, see figure \ref{fig:mag_colour}), it is interesting to investigate relationships between only the metal poor clusters in our sample. We do this by removing the two clusters with [Fe/H]$>$-0.8. For this sample of metal poor/ intermediate metallicity clusters, a significant relationship is now identified between the FUV-\textit{g} colour and the cluster core density (with N=27, $\rho_{s}$=-0.60 and P=0.001). In contrast, the relationship between FUV-\textit{g} and metallicity is relatively weak for this sample (with N=27, $\rho_{s}$=0.32 and P=0.10). Given this observed correlation between the cluster core density and FUV-\textit{g}, we investigate whether this effect can help to explain some of the scatter that is observed in the metallicity relationship. To do this, we assume a planar relationship among a cluster's FUV-\textit{g} colour, its density and its metallicity. These data are fit to the function: 
\begin{equation}
 \label{eq:fuv_g_fit}
 (\rm{FUV}-\textit{g})_{o} = \alpha\rm{[Fe/H]} + \beta\rm{log(\rho_{c})} + c
\end{equation}
From this we obtain $\alpha$=0.349 and $\beta$=-0.213. Using this relationship, we detrend our data for the effects of cluster core density and investigate whether this produces a tighter correlation between the clusters FUV-\textit{g} colour and its metallicity. This is plotted in figure \ref{fig:fit_FUVg_Z_rho0}. It can be seen that, accounting for the derived core density effect, the correlation between the metallicity and FUV-\textit{g} colour is stronger. This is confirmed by a Spearman rank test, which shows that the correlation with metallicity is stronger having removed the proposed core density effect (with N=27, $\rho_{s}$=0.57 and P=0.002). This suggests that effects due to stellar density may help to explain some of the scatter observed in the relationship between metallicity and FUV luminosity of the metal poor clusters. 

These data suggest that the cluster core density does have a significant effect on the FUV emission from these clusters. We caution that, while the above relationships are formally significant, they are based on a relatively small sample of clusters. However, we know of no reasons why selection effects should bias our results in such a way as to produce the observed relationship.

\subsection{Milky Way globular cluster data} 

\begin{figure}
 \includegraphics[width=86mm,angle=270]{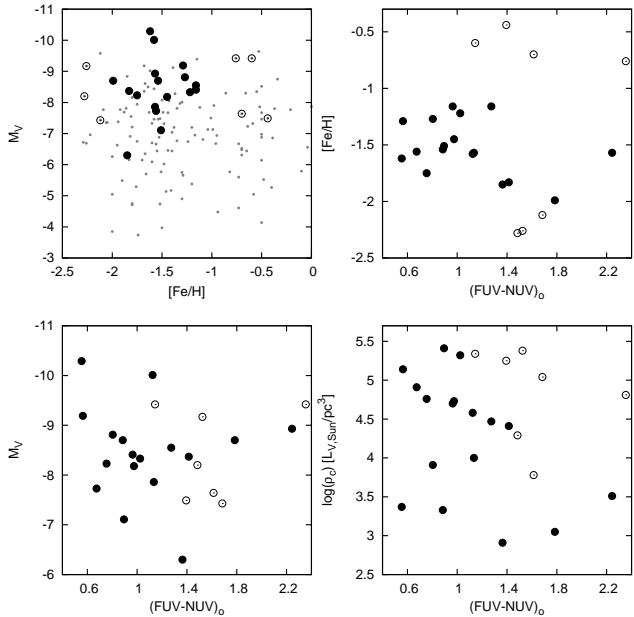}
 \caption{The UV colour of 24 of the Milky Way's GCs. UV colours are taken from \textit{ANS}, \textit{IUE} and \textit{OAO} observations as presented in \citet{Dorman95} and \citet{Sohn06}. Colours are transformed into the \textit{GALEX} filters using equation \ref{eq:AB_ST_mag}. In the top left plot, grey points indicate all Galactic GCs listed in the Harris catalogue. Open circles show all clusters with FUV and NUV photometry, filled circles highlight the clusters with intermediate metallicities. }
 \label{fig:UV_MWGC}
\end{figure}

\begin{table*}
 \footnotesize
 \centering
 \begin{minipage}{150mm}
  \caption{FUV luminosity of selected Galactic LMXBs \label{tab:FUV_MW_LMXBs}}
\end{minipage}
 \begin{minipage}{140mm}
  \centering
  \begin{tabular}{@{}lrrrr@{}}
  \hline
  \hline
  Name               & d [kpc]    & A$_{\rm{FUV}}$ & F$_{1400\rm{\AA}}$ [erg/cm$^{2}$/s/$\rm{\AA}$]  & L$_{1400\rm{\AA}}$ [10$^{32}$erg/s/$\rm{\AA}$] \\
  \hline
  Sco~X-1$^{1}$                &  2.8  & 2.430 & 2$\times10^{-13}$    &   18.0 \\
  Cyg~X-2$^{2}$                &  8.0  & 3.645 & 5$\times10^{-15}$    &   11.0 \\
  Her~X-1$^{3}$                &  5.5  & 0.405 & 7$\times10^{-14}$    &    3.7 \\
  AC211~(M15)$^{4}$            & 10.3  & 0.810 & 1.0$\times10^{-14}$  &    2.4 \\
  4U~1820-30~(NGC~6624)$^{5}$    & 7.9   & -     & 1.8$\times10^{-14}$  &    1.3 \\
  $[$ M31 GC-B022              & 780   & 0.324 & 4.0$\times10^{-17}$  &   40 $]$ \\
  $[$ M31 GC-B338              & 780   & 1.134 & 1.1$\times10^{-15}$  & 2000 $]$ \\
  $[$ M31 GC-B193              & 780   & 0.891 & 7.8$\times10^{-17}$  &  110 $]$ \\
  $[$ M31 GC-B225              & 780   & 0.810 & 4.0$\times10^{-16}$  &  550 $]$ \\
  \hline
  \end{tabular}\\
\end{minipage}
\begin{minipage}{150mm}
\footnotesize
The estimated distance to, extinction (assuming $R_{\rm{FUV}}$=8.1), FUV flux and (extinction corrected) FUV luminosity of selected Galactic LMXBs. For comparison, the integrated FUV luminosity of the the faintest and brightest M31 GCs detected by \citet{Rey07} are listed. Also listed are the FUV bright metal rich clusters in M31, which are known to host LMXBs. The Galactic LMXB data are taken from: $^{1}$ \citet{Vrtilek91}; $^{2}$ \citet{Vrtilek90}; $^{3}$ \citet{Boroson00}; $^{4}$ \citet{Dieball07}; $^{5}$ \citet{King93}.
\end{minipage}
\end{table*}

Given the relatively low number of clusters in our sample, and the observed (and expected) subtlety of any relationships, it is also interesting to consider the UV properties of the Milky Way's clusters. The distances, metallicities and structural parameters of the Milky Way GCs were taken from the Harris catalogue \citep{Harris96}. Ultraviolet observations of the Milky Way's GCs are complicated due to their large angular size, contamination (from foreground stars and background active galactic nuclei) and the high (and often variable) extinction in the direction of some of these clusters. This means that data are currently only available for a small sample of clusters. Integrated FUV and NUV photometry of 24 clusters were presented by \citet{Dorman95} using a combination of observations from the \textit{Astronomical Netherlands Satellite (ANS)}, \textit{Orbiting Astronomical Observatory (OAO)} and \textit{International Ultraviolet Explorer (IUE)}. These data were taken from table~1 of \citet[][for the \textit{IUE} data]{Dorman95} and table~6 of \citet[][which makes use of updated extinction estimates for these data]{Sohn06}. We exclude clusters whose photometry is considered unreliable by \citet{Dorman95} and the cluster NGC~6441, whose colour may be contaminated by a forground star. These data are on the STmag system and can be transformed to the \textit{GALEX} ABmag system via \citep{Sohn06}: 

\begin{equation}
 \label{eq:AB_ST_mag}
 \rm{(FUV-NUV)_{AB} = (FUV-NUV)_{STmag}+0.854}
\end{equation}

Figure \ref{fig:UV_MWGC} shows the UV colour of the Milky Way's GCs as a function metallicity, luminosity, and core density. It can be seen that the FUV-NUV colour does not get redder with increasing metallicity (with N=24, $\rho_{s}$=-0.14, P=0.51). Interestingly, there appears to be a lack of FUV bright metal poor clusters, with intermediate metallicity clusters having the bluest colours. The lack of a correlation between FUV-NUV and metallicity is in agreement with the observations of M31's clusters and adds support to the idea that metallicity has less of an influence on the UV colour than it does on the UV-optical colour. 

We also find no significant correlations between the FUV-NUV colour of the clusters and either their luminosity or core density. Limiting the sample to intermediate metallicity clusters (-2.0$<$[Fe/H]$<$-1.2, solid points in figure \ref{fig:UV_MWGC}), does identify a weak correlation between the FUV-NUV colour and core density. Unlike our M31 sample, a few of the Milky Way's low density clusters are found to have very blue FUV-NUV colours. We note that the bluest of these is $\omega$~Cen. This is the most massive cluster in our sample, so its blue colours may be suggestive of a mass effect. However, this cluster is also likely to have had a more complex star formation history than a typical cluster. As such, its UV colour may be atypical. 

The relative weakness of correlations observed in the Milky Way's clusters (compared with those observed in our M31 data) may also be due to the reliability of their UV colours. The data available for the Milky Way's clusters is less homogeneous than our \textit{GALEX} observations and their UV colours may have increased errors due to the difficulty of obtaining accurate UV photometry in the Galaxy. Indeed a previous study by \citet{Djorgovski92} used a similar \textit{ANS} and \textit{IUE} dataset to demonstrate that ``more luminous and more concentrated clusters have bluer FUV colours". It is unclear why we do not identify this relationship in figure \ref{fig:UV_MWGC}, since the data are from similar surveys. However, there are several differences between the datasets. Firstly, \citet{Djorgovski92} have a larger, but possibly more contaminated dataset, because they include eight clusters that are excluded by \citet[][and hence our study]{Dorman95} because their photometry is potentially unreliable. Slightly different UV colours are used by \citet{Djorgovski92} and they include more colours from \textit{IUE} data (which are obtained through a smaller aperture than the \textit{ANS} data, hence excluding the outer regions of some clusters but possibly suffering from less contamination from non-cluster sources). Finally, both datasets are relatively inhomogeneous, being taken from multiple studies using multiple telescopes. We believe this discrepancy highlights the difficulty of studying FUV correlations in the Milky Way's clusters (particularly considering the small sample currently available). We do not attempt to draw any new conclusions from these Milky Way data, other than to suggest that it does not disagree with our conclusions based on M31's clusters. The need for higher quality UV photometry of the Galactic GCs is clear, and may become available from analysis of future/archived \textit{SWIFT/GALEX} observations (although such work will still have to contend with contamination and extinction issues). 

Data for a larger sample of the Milky Way's clusters are available from studying optical CMDs. Using HST observations of 54 clusters, \citet{Recio_Blanco06} have identified that the more massive globular clusters do tend to host bluer HB stars (consistent with our findings in M31). This relationship is particularly strong for intermediate metallicity clusters (where the influence of metallicity is relatively small; see their figure 4). A relationship has also been identified between the length of a cluster's `blue tail' and its core density \citep{Fusi_Pecci93,Buonanno97}. Both of these findings are in agreement with our conclusions based on integrated UV observations of M31's clusters. Interestingly, such a core density relationship was \textit{not} found in the study of \citet{Recio_Blanco06}. The reason for this discrepancy is not clear. However, it possibly highlights the difficulty in accurately studying the blue extremes of a clusters HB population using optical CMDs.

\subsection{Low mass X-ray binaries and other FUV-bright sources} 

It is generally assumed that, if a cluster hosts a significant population of BHB, EHB or BHk stars, then these will dominate the integrated FUV luminosity of the cluster. This is because most other stars in a cluster will be too cool to emit significantly in the FUV. However, there are other FUV bright sources in GCs. We consider the influence of these on the clusters integrated luminosity below. 

Low mass X-ray binaries (LMXBs) are found in the cores of many GCs. These objects can have very high X-ray luminosities which can irradiate the systems accretion disks and$/$or donor stars and heat them to very high temperatures. The result of these large, hot, accretion disks is that LMXBs can be very bright in the FUV. Indeed, the LMXB 4U~1820-30 in the Galactic GC NGC~6624 dominates the total FUV luminosity of the cluster \citep{King93}. Even in the GC M15, which hosts a large population of FUV bright HB stars, the LMXB AC211 is the brightest FUV source in the cluster core \citep{Dieball07}. In table \ref{tab:FUV_MW_LMXBs} we list the FUV luminosity of these two Galactic GC LMXBs and three other Galactic LMXBs. The fluxes of these LMXBs are highly variable. This table therefore provides only an estimate of the highest FUV fluxes reached during these observations. It can be seen that the bright LMXBs Sco~X-1 and Cyg~X-2 are also very bright FUV sources. However, in the context of GCs, it should be noted that GCs are unlikely to host more than one or two LMXBs in outburst. This means that, for clusters with a significant BHB population, a single bright LMXB is unlikely to contribute more than $\sim10\%$ to the integrated FUV luminosity of the clusters. However, for metal rich GCs (which generally lack BHB stars), LMXBs could dominate the emission. This effect will be enhanced by that fact that metal rich clusters preferentially form LMXBs \citep[see e.g.][for the first clear evidence of this effect]{Kundu02}. It can be seen from table \ref{tab:FUV_MW_LMXBs} that a single LMXB is unlikely to explain the FUV bright metal rich clusters identified by \citet{Rey07}. However, it is interesting to note that two of the three metal rich clusters they identified do host LMXBs (B225 and B193). In particular, the cluster B225 hosts an LMXB with $L_{x}>10^{38}$ergs/s \citep{Trudolyubov04}. Such a source will be very bright in the FUV. 

Single `UV bright' stars, such as post-AGB stars, can also contribute a significant fraction of the integrated luminosity of a cluster. However, clusters are also unlikely to host many of these sources due to their short lifetimes. It is thought that the contribution from such stars, to the integrated luminosity of a cluster, is unlikely to be more than 15$\%$ \citep{Moehler01}. 

In addition to these objects, there are several other hot populations in GCs. Cataclysmic variables (CVs; white dwarfs accreting from Roche lobe overflowing companions) are relatively common in GCs. Because of their decreased bolometric luminosities, CVs are generally fainter than LMXBs in the FUV. However, they are more numerous and can reach FUV luminosities similar to a typical BHB star. Single white dwarf stars also emit strongly in the FUV. The average luminosity of both CVs and white dwarfs are thought to be too low to make a significant contribution to the integrated FUV luminosity of a cluster. The other notable FUV sources in GCs are blue straggler stars. These are very bright, blue objects and can also reach FUV luminosities similar to BHB stars, although blue stragglers this bright are likely to be rare. These stars are also unlikely to make a significant contribution to the integrated FUV luminosity of a cluster with a large population of blue HB stars. However, they may become increasingly important for clusters with redder HB populations. 

\section{Conclusions}

We have considered the UV properties of M31's GCs. The previously identified relationships between the metallicity of a cluster and both its FUV-\textit{g} and NUV-\textit{g} colours are confirmed. Considering M31's metal poor/ intermediate metallicity clusters, we identify a relationship between the clusters core densities and their FUV-\textit{g} colours. This suggests that stellar density may influence the HB stars in these clusters. We demonstrate that the effects of stellar density on the properties of HB stars in a cluster may help to explain some of the scatter in the observed metallicity relationships. We also identify a weak anti-correlation between the FUV-NUV colour of a cluster and its metallicity. While this relationship is not very significant, it is in the opposite direction to the general relationship between metallicity and UV-optical colours. This suggests that the metallicity of a cluster may have little effect on the blue extremes of the HB distribution. 

These data show a significant relationship between the FUV-NUV colours of these clusters and their core densities. We interpret this result in the context of a population of EHB stars, the production of which is enhanced in dense stellar environments. This may be due to enhanced mass loss or helium enhancement, in close encounters, or dynamical formation of tight binary systems (in which mass loss can occur, forming the EHB stars). We also find that more massive clusters tend to have bluer UV colours. This may be related to the same mechanisms suggested for the core density relationship, or it may point to an effect due to self enrichment. These trends are not confirmed in a smaller sample of integrated UV colours of the Milky Way's GCs. However, we note that this data is less homogeneous (and may suffer from more contamination) than our M31 GC data and that other studies of the Milky Way's clusters have suggested such trends (based on both UV and optical observations). It should also be noted that, if such relationships are present, they are expected to be relatively weak. 

The relationship between cluster density and the population of EHB stars is potentially important in understanding the formation of these stars and the distribution of HB stars in clusters. Also, if a cluster's density does have a significant influence on its FUV emission, then it is important to consider this when using FUV observations to estimate the ages of globular clusters. Currently, such work does not take density effects into account. 

We caution that, while these data suggest interesting correlations, the relationships observed are for a relatively small sample of M31's clusters. Obtaining new measurements of the structural parameters, metallicities and reddenings for more of M31's clusters would allow us to extend this work to a larger sample of clusters. Obtaining deeper UV data would also enable us to investigate similar relationships in the metal rich clusters in M31. However, there are no known reasons why the data currently available should bias us towards the observed relationships. 

The FUV emission from other sources in a GC are also considered. These sources are unlikely to dominate the integrated FUV emission of a cluster if it contains a significant population of BHB, EHB or BHk stars (such as the FUV bright clusters considered in this study). However, other stellar populations, in particular LMXBs and blue stragglers, may contribute a significant fraction to the integrated FUV luminosity in clusters with only a red HB population. 

\section*{Acknowledgements}

We would like to thank Steve Zepf and Arunav Kundu for very helpful discussions related to this work. We would also like to thank the reviewer of this paper for their helpful comments. 

\bibliographystyle{mn2e}
\bibliography{bibliography_etal}

\label{lastpage}

\end{document}